\input epsf.sty
\documentclass{aa}
\begin{document}

\def\Agata{R\' o\. za\' nska~}
\def\cs{c_{\rm s}}
\def\Rtr{R_{\rm tr}}
\def\rtr{r_{\rm tr}}
\def\Ti{T_{\rm i}}
\def\Te{T_{\rm e}}
\def\Ts{T_{\rm s}}
\def\Tv{T_{\rm v}}
\def\Fadv{F_{\rm adv}}
\def\Fsoft{F_{\rm soft}}
\def\Fcond{F_{\rm cond}}
\def\mdotz{\dot m_{\rm z}}
\def\mH{m_{\rm H}}
\def\me{m_{\rm e}}
\def\OmegaK{\Omega_{\rm K}}
\def\Prad{P_{\rm rad}}
\def\Ptot{P_{\rm tot}}
\def\Pgas{P_{\rm gas}}
\def\Pbase{P_{\rm base}}
\def\Revac{R_{\rm evac}}
\def\Rec10{R_{\rm evac}}
\def\Revap{R_{\rm evap}}
\def\tauevap{\tau_{\rm evap}}
\def\tauv{\tau_{\rm visc}}
\def\kappaes{\kappa_{\rm es}}
\def\kappao{\kappa_{\rm o}}
\def\alphac{\alpha}
\def\Qi{Q^{+}_{\rm i}}
\def\Q+e{Q^{+}_{\rm e}}
\def\Ftot{F_{\rm tot}}
\def\Fdisk{F_{\rm disk}}
\def\fdisk{f_{\rm disk}}
\def\fcor{f_{\rm cor}}
\def\Fcor{F_{\rm cor}}
\def\Fsoft{F_{\rm soft}}
\def\Fd{F_{\rm d}}
\def\RS{R_{\rm Schw}}
\def\rmax{r_{\rm max}}
\def\nuei{\nu_{\rm ei}}
\def\Fcool{F_{\rm cool}}
\def\taues{\tau_{\rm es}}
\def\MSun{{\rm M}_{\odot}}
\def\Hc{H_{\rm c}}
\def\const{{\rm const}}
\def\lambdaF{\lambda_{\rm F}}
\def\mcor{\dot m_{\rm cor}}
\def\mc26{\dot M_{\rm cor26}}
\def\mdotevapbranch{\dot m_{\rm evap.branch}}
\def\Ka{K$_{\alpha}$~}
   \thesaurus{03      %extragalactic
              (02.18.7;  % Radiative transfer,
	       02.01.2;  % Accretion disks,
               11.01.2;  % Galaxies:active,
	       11.17.3;  % quasars:general
	       11.19.1;  % Galaxies:Seyfert
	       13.25.2;  % X-rays:galaxies.
               13.25.5)}  % X-rays:stars

\title{Reprocessing of X-rays in AGN. I. Plane parallel geometry -- test of 
pressure equilibrium}
\titlerunning{X-ray reprocessing in AGN}

   \author{A.-M. Dumont,
          \inst{1}
	B. Czerny,
	   \inst{2}
          S. Collin\inst{1}
	\and P.T. \. Zycki\inst{2}
%\fnmsep\thanks{Just to show the usage
%          of the elements in the author field}
          }

   \offprints{A.-M. Dumont}

   \institute{$^1$Observatoire de Paris-Meudon, DAEC, Meudon, France \\
   $^2$Copernicus Astronomical Center, Bartycka 18, 00-716 Warsaw, Poland \\
%             \thanks{The university of heaven temporarily does not
%                     accept e-mails}
             }

   \date{Received ...; accepted ...}

   \maketitle

   \begin{abstract}

We present a model of the vertical stratification and the spectra of an
irradiated medium under  the assumption of  constant pressure. Such a solution 
has properties intermediate between constant density models and 
hydrostatic equilibrium models, and it may represent a flattened configuration 
of gas clumps accreting
onto the central black hole. Such a medium develops a
hot skin, thicker than
hydrostatic models, but thinner than constant density models, under comparable
irradiation. The range of theoretical values of the $\alpha_{\rm ox}$ index is 
comparable to those from hydrostatic models and both are close to the observed 
values for Seyfert galaxies but lower than in quasars. The amount of 
X-ray Compton reflection is consistent with the observed range. The 
characteristic property of the model is  a frequently multicomponent iron 
\Ka line.  
  
\keywords{Radiative transfer, Accretion disks, Galaxies:active, Galaxies:Seyfert, X-ray:stars, X-rays:galaxies
               }
   \end{abstract}

%
%________________________________________________________________

\section{Introduction}

The intense production of hard X-ray emission is a basic property of all
active galactic nuclei (AGN), although the fraction of this emission in
comparison to the total bolometric luminosity is different in various types of
AGN. A fraction of this radiation illuminates the relatively cold material
present there and provides a reflection component apparent in X-ray spectra
(Pounds et al.\ 1990; for a review see Mushotzky et al. 1993). 
The component consists of an iron \Ka line,
usually at 6.4 keV appropriate for only weakly ionized iron,
an accompanying edge, and a Compton ``hump'' peaking at about 
30 keV.   

Above 1 keV the  
spectral index is equal to 0.7 on average, and in a majority of objects the continuum 
below 1 keV displays
 an excess when compared to the extrapolation of the
continuum in the 1-10 keV range (the so-called ``soft X-ray excess").  Recent 
observations performed with {\it Chandra\/} show that the strength of 
the excess has probably been overestimated in the past. This soft X-ray excess
may, or may not, be related to the X-ray reprocessing (e.g. Czerny \&
\. Zycki 1994, Leighly 1999).

Above 1 keV the spectrum can 
be decomposed into an
underlying power law (with a typical spectral index $\Gamma \sim 1.9$) and a 
reflection component. The strength of the reflection  component correlates
strongly with the slope of the power law,  and this effect is universal
for all accreting black holes, i.e. AGN and galactic black holes 
(Ueda et al. 1994, Magdziarz et al. 1998, Done \& \. Zycki 1999,
Zdziarski et al. 1999, Gilfanov et al. 1999, 
Revnivtsev et al. 2001). Frequency-resolved spectroscopy puts
additional constraints on the formation of the reflection component (Gilfanov
et al. 2000, Revnivtsev et al. 1999, \. Zycki 
2002) but this powerful technique was not applied so far to AGN observations.

The iron \Ka line displays an asymmetrical profile 
with  an extended red wing and a much smaller blue wing
(Nandra et al.\ 1997). 
The presence of the broad red wing in the Seyfert composite
spectrum has been recently questioned 
(Lubi\' nski \& Zdziarski, 2001). However, in the two best studied 
objects, MCG-6-30-15 and 
NGC 3516, the line wing extends down to 4 keV (see Tanaka et al.\ 1995, Iwasawa
et al.\ 1996, Lubi\' nski \& Zdziarski 2001 for MCG-6-30-15, and Nandra et al.\ 
1999 for NGC 3516). Recent high resolution {\it Chandra\/} data show a neutral 
very narrow iron line in most objects (e.g.\ Yaqoob et al.\ 2001a; 
Sambruna et al.\ 2001; Kaspi et al.\ 2001). These results most 
probably mean that the line is multicomponent, as actually shown by 
Yaqoob et al.\ (2001b).
 The equivalent 
width of the line seems also 
to correlate with the spectral slope of the power law 
(Lubi\' nski \& Zdziarski, 2001). 

The iron line and the reflection hump have been now
extensively used for the diagnosis of the geometry of the accretion flow 
pattern close to the black hole (e.g.\ Fabian et al.\ 2000, 
Yu \& Lu 2000, Ruszkowski et al.\ 2000, 
Abrassart \& Dumont 1998, 2000, Karas et al.\ 2000,
\. Zycki \& \Agata 2001, and references therein).

Recent intensive monitoring campaigns brought, however, more confusion 
 than
enlightening to the emerging picture. According to predictions based on 
a simple analysis, a variable X-ray emission should give in response
a significant variability in the optical/UV band due to the absorption of a 
fraction of incident X-rays. Also both the reflected
X-ray continuum and the iron line should vary in concert, though with 
a small delay
(e.g. Rokaki et al. 1993, Clavel et al.\ 1992, Berkley,
Kazanas \& Ozik 2000, Kazanas \& Nayakshin 2001). Recent results of 
careful monitoring, however, showed that the level of the optical/UV 
response is 
surprisingly different in various sources, the correlation between the
optical/UV and X-rays is far from perfect
or is actually absent, the iron line total flux responds only weakly or not at all
to variable X-rays, and the level of reflection in continuum is not
positively correlated with the strength of the line. 

For instance, NGC 7469 shows 
variations of the UV and X-ray fluxes with similar large
amplitudes and a time delay of the order of 2 days, the UV leading the 
X-rays in the flux peaks, but the X-ray flux
 displays in addition short term variations not seen in UV (Nandra et al. 
1998). The UV flux correlates with $\Gamma$ (Nandra et al, 2000).
For NGC 3516 there is no clear correlation between UV and X
(Edelson et al.\ 2000). For NGC 4051 no optical variations were observed when 
strong X-ray variations were seen (Done et al.\ 1990). 
 The UV flux variations drive the  optical flux
variations with a time delay $\le 0.2$d for NGC 
4151 (Edelson et al.\ 1996), $\le 0.15$d for NGC 3516 (Edelson et al. 
2000), and for NGC 7469 a delay of the order a day which increases with 
the wavelength (Wanders et al.\ 1997, Collier et al.\ 1998). 
Soft X-ray variations are 
generally larger than hard X-ray ones, but 
this is not always the case (Nandra et al.\ 1997). For NGC 3516, there were 
strongly correlated variations with no measurable lag ($\le$ 0.15d, Edelson et al.\ 2000). 

The centroid energy, the intensity, and the equivalent width
of the Fe \Ka line, are variable on short time scales (down 
to ksec). The correlations between the continuum and the Fe \Ka line
 have been intensively looked for in a few objects with ASCA and RXTE, and now 
with {\it Chandra\/}, revealing a 
complex behaviour of the line versus the underlying continuum. 
Iwasawa et al.\ (1996) have observed rapid variations of the Iron line 
profile and intensity in 1 to 10 ksec in MCG-6-30-15 on the basis of an ASCA 
study. But this result has been recently questioned 
by Lee et al.\ (1999) and Chiang et al.\ (2000) who have studied with RXTE 
the spectral 
variability of MCG-6-30-15 and of NGC 5548, and have shown that the Fe \Ka
line flux is constant over time 
scales of 50 to 500 ksec, while the underlying continuum displays large 
flux and spectral variations, and the profile of the line is variable! The 
same result seems to be observed in NGC 3615. Reynolds (2000) extended the 
work of Lee et al.\ (1999) on MCG-6-30-15, excluding a time delay between the 
line and the continuum 
in the range 0.5 to 50 ksec, and suggesting that the line flux remains 
constant on 
timescales between 0.5 and 500 ksec. From an analysis of the ASCA archive 
data of a sample of Seyfert nuclei, Weaver et al. (2001) found that 
in most  cases changes in the line do not appear to track changes
in the continuum.

These problems indicate, that either the geometry of the accretion flow
envisioned as an X-ray lamp above the flat disk plane is not correct, or the
response of the material is not as expected from simple estimates, or both.

In the present paper we concentrate on the second problem, i.e.\ on 
an accurate description of the X-ray reprocessing by an optically thick material.
Since we do not want to restrict ourselves
to a particular model of accretion flow we discuss two basic general cases:
we consider a medium at constant pressure (with the density gradient 
determined by the
temperature profile and the radiation plus gas pressure gradient) and we compare this 
situation to the case of a medium 
at constant density. We assume a plane parallel geometry. 
In the next paper we will explore the influence of the 
geometry, with particular attention given to a quasi-spherical case.

The radiative transfer 
computations are done with the coupled codes {\sc titan} 
and {\sc noar} described
in detail by Dumont et al. (2000; hereafter DAC). We present a
sequence of models characterized by a broad range of values of the 
ionization parameter defined at the surface. We also study the effects of
other parameters involved, like the shape of the incident radiation flux.

We show the broad band spectra,  characterize their overall properties
by several parameters frequently measured observationally, like the 
$\alpha _{\rm ox}$
index and the amount of reflection as seen in XSPEC analysis, and finally we
discuss the results comparing these quantities to typical parameters in
various types of AGN.

\section{Radiation transfer}
\label{sec:rad}

The code  {\sc titan} was developed for the purpose of radiative transfer
computations in an  optically thick irradiated medium. Basically similar codes 
were 
developed by  Ross \& Fabian (1993), Nayakshin et al. (2000;
hereafter NKK),
Madej \& R\. o\. za\' nska (2000).
Other codes either do not cover the optical/UV energy range 
(\. Zycki et al.\ 1994), or do not include hard X-ray 
irradiation (Hubeny et al.\ 2001), or
they include hydrogen and helium only  
(Sincell \& Krolik 1997).

The code {\sc titan}, the code of Ross \& Fabian (1993) and the NKK code based
 on {\sc xstar} 
include the non-LTE effects while Madej \& R\. o\. za\' nska (2000) 
still work in the LTE approximation. Both Ross \& Fabian (1993) and NKK 
include the effect of Comptonization on the spectral distribution
of photons while  {\sc titan} considers only the effect of Compton 
heating/cooling in energy balance. In order to obtain the Comptonized spectrum,
 the transfer
code  {\sc titan} has to be supplemented with the Monte Carlo 
code {\sc noar} of DAC.
The code of Ross \& Fabian (1993), adopted by Ballantyne, 
Ross \& Fabian (2001), contains only a small 
number
of basis atomic lines. The major 
difference of  {\sc titan} with other codes (except
that of Hubeny) is that it solves the line transfer instead of using the
escape probability approximation.

The detailed description of the code {\sc titan} and exemplary results for the
temperature profile, ionization stage profile and the spectra in
case of constant density of the medium are given by DAC and in 
Dumont \& Collin (2001).
The code was already used to reproduce the mean quasar spectra 
within the frame of the cloud model (Czerny \& Dumont 1998), to model
the iron line (after supplementing code results with {\sc noar}: 
Abrassart 2000, Abrassart \& Dumont 1998, 2000,
Karas
et al.\ 2000) and to model the reflection by an accretion disk (after
coupling with the computations of the hydrostatic 
equilibrium: R\' o\. za\' nska et al.\ 2002).

\section{Results}

Since the dynamics of the formation of the hot and cold medium close to a 
black hole is not known it is necessary to reduce the problem to the
consideration of 
 a few basic 
possibilities for the description of the irradiated medium. 

If the cold medium consists of clumps of cold material formed through
a violent disruption of the disk accretion flow, the clouds may not reach
an equilibrium and may be represented as a constant density medium. However,
if the clouds are illuminated and can survive long enough, the medium will 
approach 
pressure equilibrium and an assumption of constant pressure may be a
better representation of the medium, with the density gradient maintained
due to the temperature and radiation plus gas pressure gradient. 
Finally, if the disk is not disrupted and it
extends down to the marginally stable orbit (and possibly below it) the 
irradiated
upper layers of the disk will be mostly in hydrostatic equilibrium. The last
possibility is characterized by the steepest density gradient.

In the present paper we concentrate on the intermediate case, i.e. the
medium in pressure equilibrium. We assume a plane-parallel geometry, and
the medium is irradiated by an isotropic incident flux. The irradiated
layers expand, and in the equilibrium situation a constant pressure is 
maintained
throughout the entire zone. This pressure includes both the gas pressure and
the radiation pressure.

\subsection{Constant pressure models}

We show first a set of models parameterized by the value of the ionization parameter
$\xi$ at the surface of the irradiated medium ($z = 0$). This parameter is defined as
\begin{equation}
\xi = {L_{\rm bol} \over n(z=0) R^2},
\label{eq:xi}
\end{equation}
where $L_{\rm bol}$ is the bolometric luminosity of the source located at a 
distance $R$ and $n(z=0)$ is the number density at the surface of the medium. 
In our computations the density was fixed
at the value $n(0) = 10^{11} {\rm cm}^{-3}$.

Radiation pressure is locally calculated from
\begin{equation}
P_{\rm rad}= {4 \pi \over 3 c} {\int J_{\rm \nu} d \nu},
\label{eq:radpress}
\end{equation} 
where $ J_{\rm \nu}$ is the mean intensity at a given depth computed at each step by the code.

The value of the total 
pressure (i.e. radiation + gas pressure) is calculated when the temperature of
the plasma has been  determined by the thermal balance equation, and this 
value is kept fixed within the medium
thus determining the density profile. The model is iterated until  
convergence of the flux (and therefore of the radiation pressure) as in DAC.

Our basic sequence of models is specified as follows.

The shape of the incident radiation was assumed to be a power law, with the 
energy index $\alpha = 1.0$, extending from 0.1 eV to 100 keV.

The total hydrogen column in these computations is fixed at 
$ 3 \times 10^{25} {\rm cm}^{-2}$ ($\tau_{\rm Th} \approx 24$).

The set of models with $\xi = 300, 10^3, 10^4, 10^5$ and $10^6$ illustrates
the effect of increasing the incident flux from 
$2.7 \times 10^{12}$ erg s$^{-1}$ cm$^{-2}$  to 
$8 \times 10^{15}$ erg s$^{-1}$ cm$^{-2}$ correspondingly.

In all cases the irradiation results in the formation of a hot surface zone 
with  an almost constant temperature. Initially the temperature decreases 
inwards rather slowly, with a 
rapid drop  by 2 or 3 orders of magnitude below  a
certain depth.

This rapid transition reflects the ionization instability which develops in
an irradiated medium (Krolik et al. 1981), and it leads to a
temperature
discontinuity if the medium is supposed to be in hydrostatic equilibrium
(Begelman et al. 1983, \Agata \& Czerny 1996, Ko \& Kallman 1994,
NKK). Unique, continuous solutions are obtained only if energy transfer by
conduction is
included, in addition to radiative heating and cooling (Begelman \& McKee 1990,
\Agata 1999). However, at present no computations include both the full 
radiative transfer and the conduction term; either radiative transfer is
 replaced by a parametric description of heating and cooling (\Agata 1999)
or the radiative transfer is computed carefully but the conduction is 
neglected (NKK, Ballantyne et al.\ 2001, \Agata et al.\ 2002 and other papers 
presenting spectra). In the second approach the instability is
dealt with, either by adopting a discontinuity at an arbitrary position 
somewhere inside the multiple solution zone, or the smooth solution is
found at the expense of a small pressure inversion in the multiple solution 
zone (see detailed
discussion of this issue by \Agata et al.\ 2002). The conditions for the
existence of multiple solutions under various approaches to line heating
and cooling will be discussed by Coup\' e et al.\ (in preparation). 
In the present paper we adopt the continuous solution approach, as 
in \Agata et al.\ (2002). 

\begin{figure*}
 \parbox{\textwidth}{
   \parbox{0.5\textwidth}{\epsfxsize = 0.5\textwidth \epsfbox[18 200 592 718]{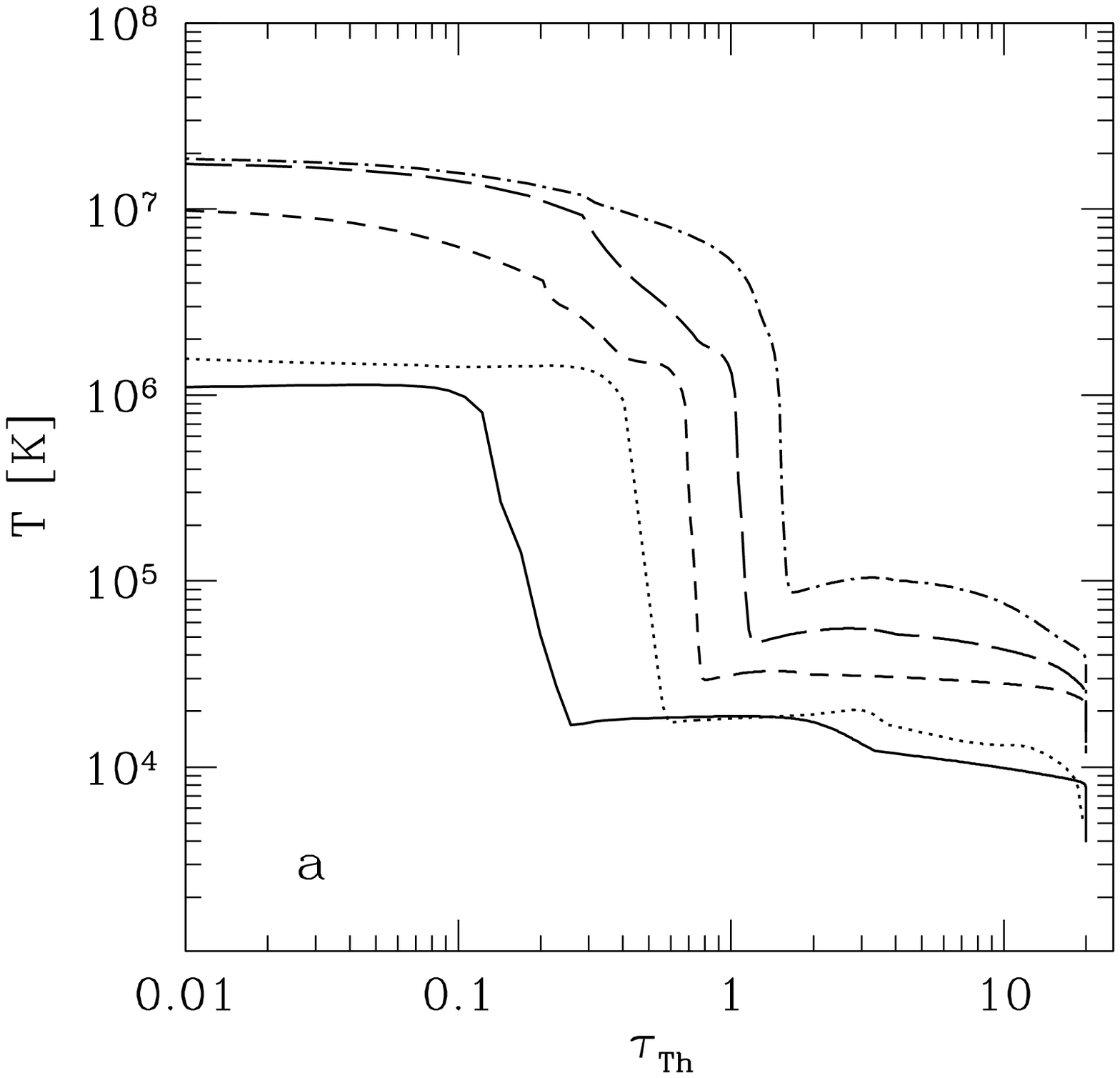}
    }
   \parbox{0.5\textwidth}{\epsfxsize = 0.5\textwidth \epsfbox[18 200 592 718]{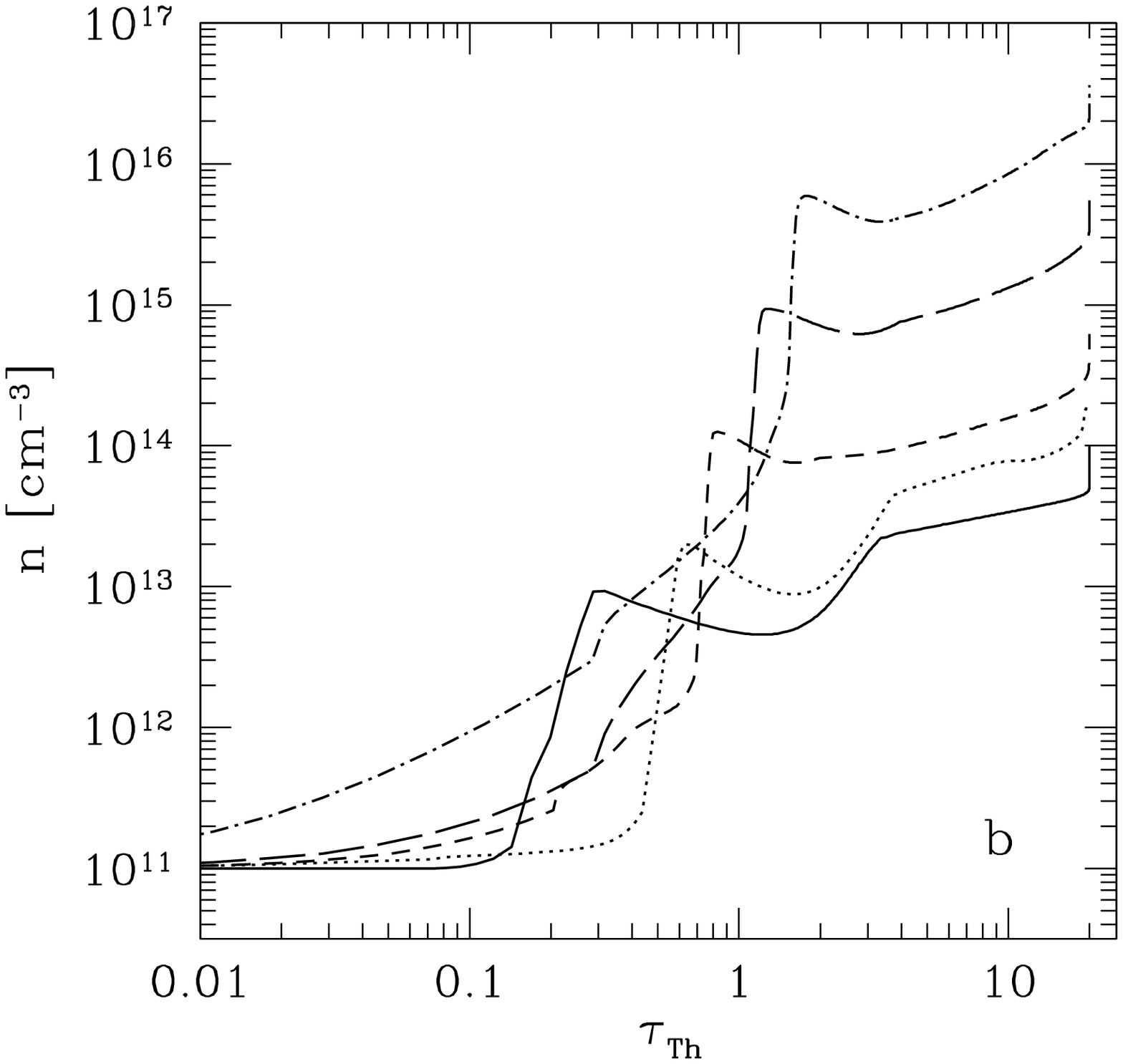}
    }
 }
 \parbox{\textwidth}{
   \parbox{0.5\textwidth}{\epsfxsize = 0.5\textwidth \epsfbox[18 200 592 718]{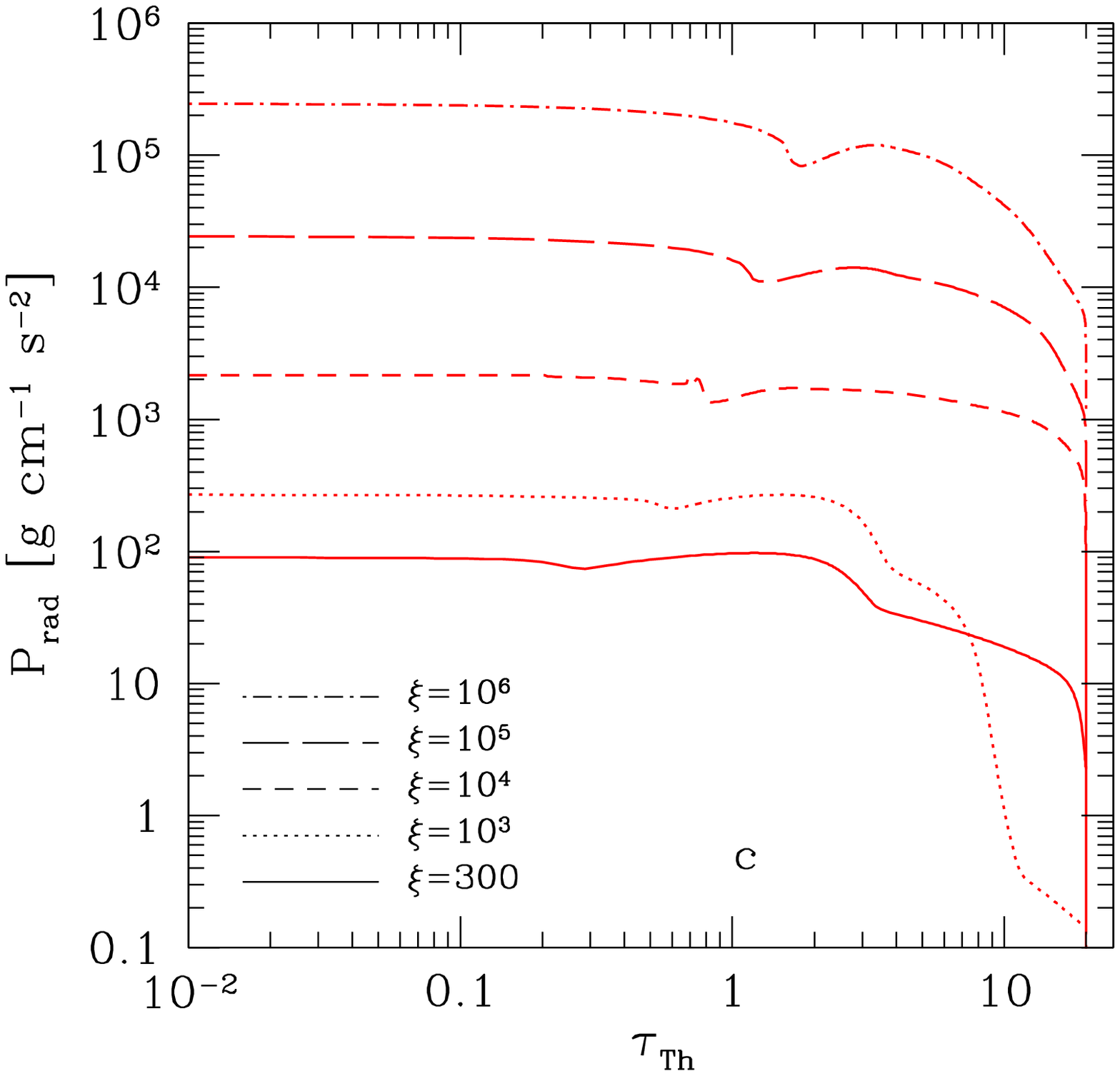}
    }
   \parbox{0.5\textwidth}{\epsfxsize = 0.5\textwidth \epsfbox[18 200 592 718]{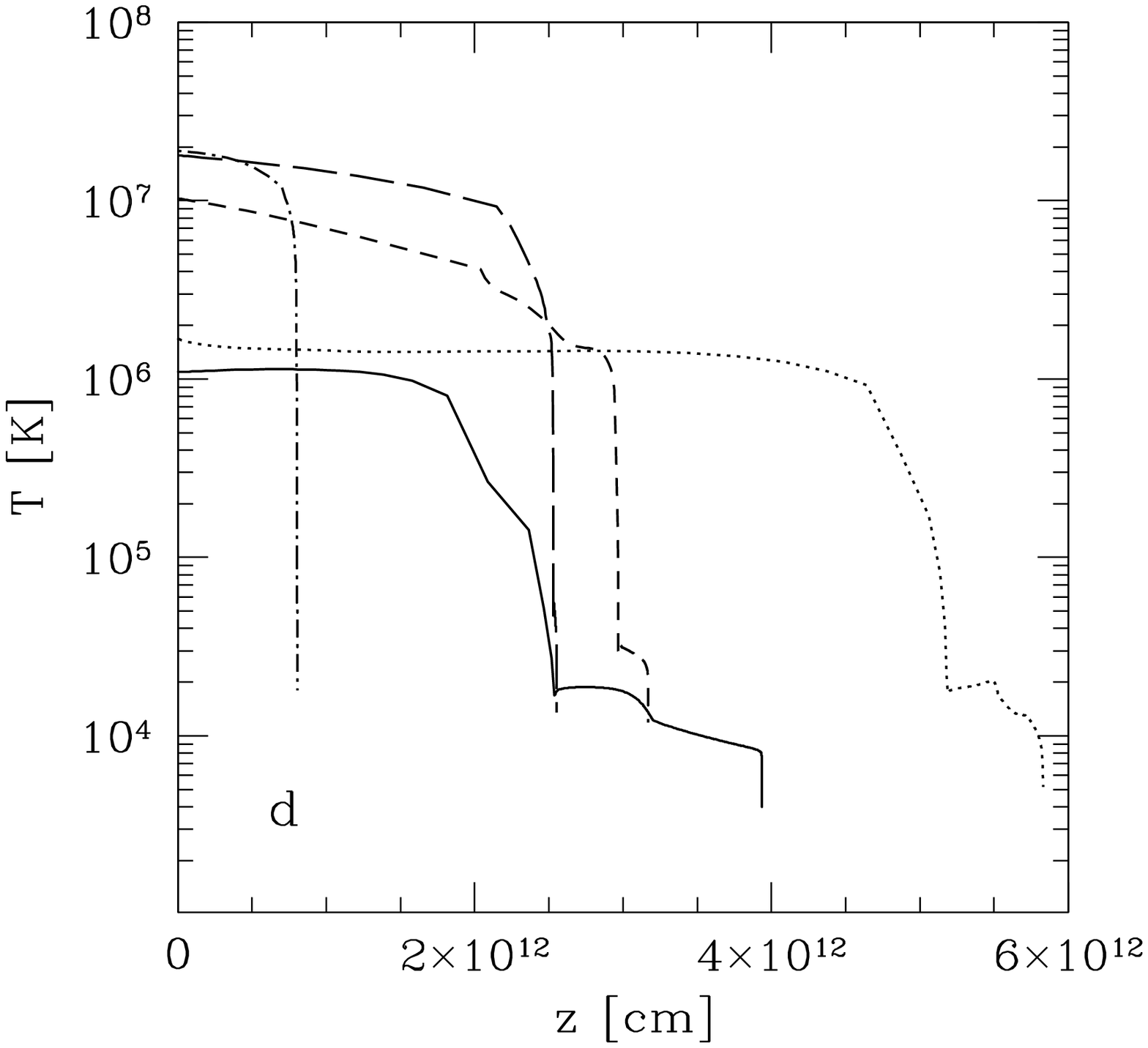}
    }
 }
\caption{The temperature (panel a), density (panel b) and pressure (panel c) 
profiles as a function of the optical depth in 
Thomson units, in  a constant pressure medium, for an ionization parameter 
at the surface $\xi = 300, 10^3, 10^4, 10^5$ and $10^6$ (continuous, dotted, 
short-dashed, long-dashed and dash-dot line, correspondingly); the density at the 
surface is $n(0)=10^{11}{\rm cm}^{-3}$, the column density 
is $N_{\rm H} = 3 \times 10^{25}{\rm cm}^{-2}$. The temperature profile as a function of the distance from the surface is shown in panel d.}
\label{fig:prof1}
\end{figure*}

We see in Fig. \ref{fig:prof1} (panel a) that both the surface temperature and the
optical thickness of the hot layer depend strongly on the incident flux. Only 
the outermost layers of $\tau_{\rm es} \sim 0.3$, and only at high irradiation 
flux, reach the Inverse Compton temperature, $T_{\rm IC}$, corresponding to the 
incident
spectrum ($1.8 \times 10^7$ K),  mostly because of the contribution coming from the thermalized  
back side radiation. For $\xi \leq 10^3$ the surface
temperature is much lower than $T_{\rm IC}$ since the ionization of the upper 
layers is
not high enough to make the Compton heating/cooling to dominate. 
The optical depth of the hot zone ranges from 
$\tau_{\rm h} \sim 0.6$ in the low flux case to $\tau_{\rm h} \sim 2$. 
Still lower ionization flux would lead to still lower optical depth of the 
hot skin and even lower surface temperature value. We do not show an example
since the temperature of the back side of the slab approaches low values
($T \le 10^4 $ K) for which the code is not adapted.

As we assume a constant pressure the density profile results from the
computations (see Fig. \ref{fig:prof1}, panel b). The density contrast between the
illuminated and back sides of the slab is very high since it reflects
the temperature gradient as well as the considerable contribution from the
radiation pressure which, in the case of high $\xi$,  dominates the gas 
pressure by several 
orders of magnitudes in the hot layers (see Fig. \ref{fig:prof1}, panel c).

This density gradient means that, while the
cold region dominates the medium in terms of optical depth, the opposite is 
true if the spatial extension of the medium is considered. For example, 
for $\xi = 10^3$ the hot
region is $5 \times 10^{13}$ cm deep and the cold region is $ 10^{13}$ cm, 
while for $\xi = 10^6$ the hot region is $8 \times 10^{11}$ cm 
and the cold region only $4 \times 10^9$ cm! Of course the cold medium can be 
geometrically and optically thicker but we do not consider larger values of 
the column density since that
would not change the reflected spectra. Therefore, apart from the temperature
plots as function of the optical depth we also plot the temperature against the
true distance from the irradiated surface (see Fig.~\ref{fig:prof1}, panel d).

The large relative geometrical size of the heated layer has an important 
consequence 
in the case of the special geometry of a quasi-spherical distribution of 
clouds irradiated by the central 
source, as discussed by Collin-Souffrin et al.\ (1996). If the clouds have 
time to reach pressure equilibrium their irradiated parts would expand
enormously and  
the medium will actually look more like clouds embedded in a warm bath, 
particularly for high irradiation flux. This will be particularly important 
when
we will discuss the spherical geometry (Dumont et al., in preparation).

\begin{figure}
\epsfxsize=8.8cm \epsfbox[18 200 592 700]{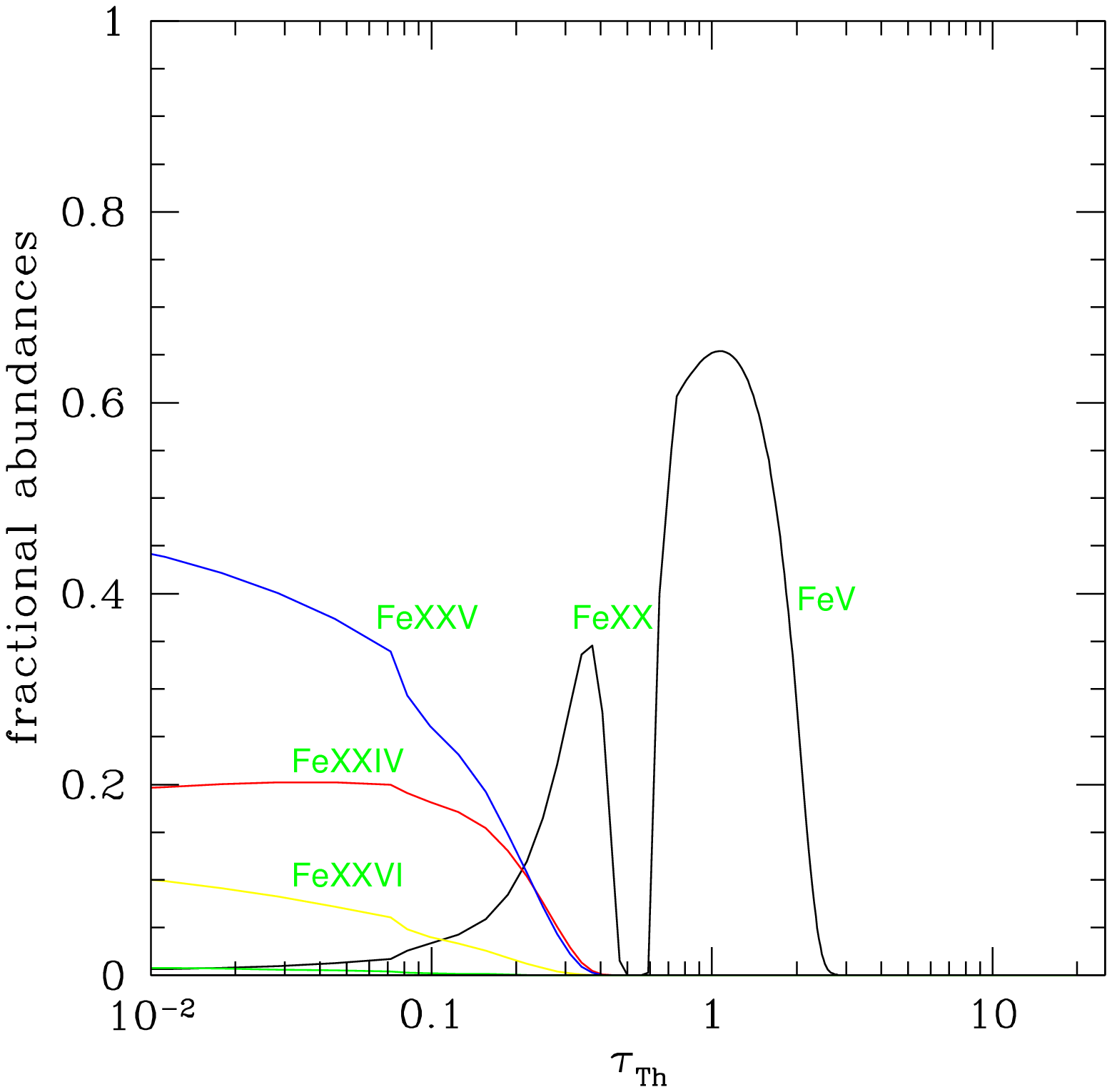}
\epsfxsize=8.8cm \epsfbox[18 200 592 700]{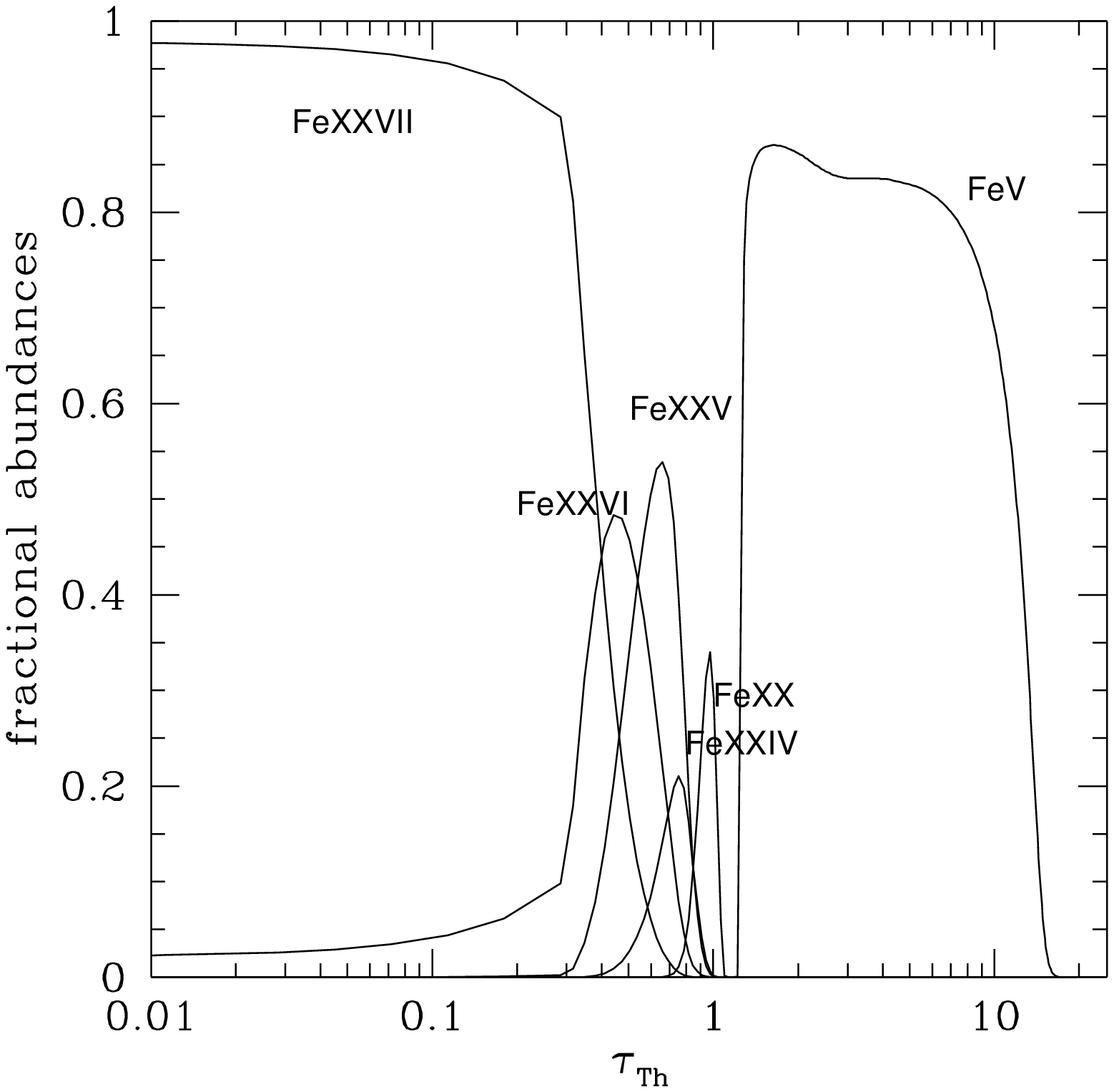}
\caption{The population of iron ions as a function of optical depth 
in a constant pressure medium, for $\xi =
10^3$ (upper panel) and $10^5$ (lower panel); $n(0)=10^{11}{\rm cm}^{-3}, N_{\rm H} = 3 \times 10^{25}{\rm cm}^{-2}$. }
\label{fig:ion1}
\end{figure}

As we mentioned earlier, the ionization of iron is complete only close
to the surface, and only in the
case of the highest irradiation flux. Even for $\xi = 10^6$ a large 
part of the
hot layer is only partially ionized and populated with hydrogen and helium-like
iron (see Fig.~\ref{fig:ion1}). The solution for $\xi = 10^3$ is still mostly
populated by FeXXV. 
Only in the case of the lowest value of the ionization parameter
the contribution from highly ionized iron becomes negligible and the line is
dominated by the 6.4 keV component. In all other cases the iron line consists
always of a number of components at various energies which reflect the thermal
stratification of the medium.                     

\begin{figure*}
\epsfxsize=\textwidth \epsfbox[30 200 600 700]{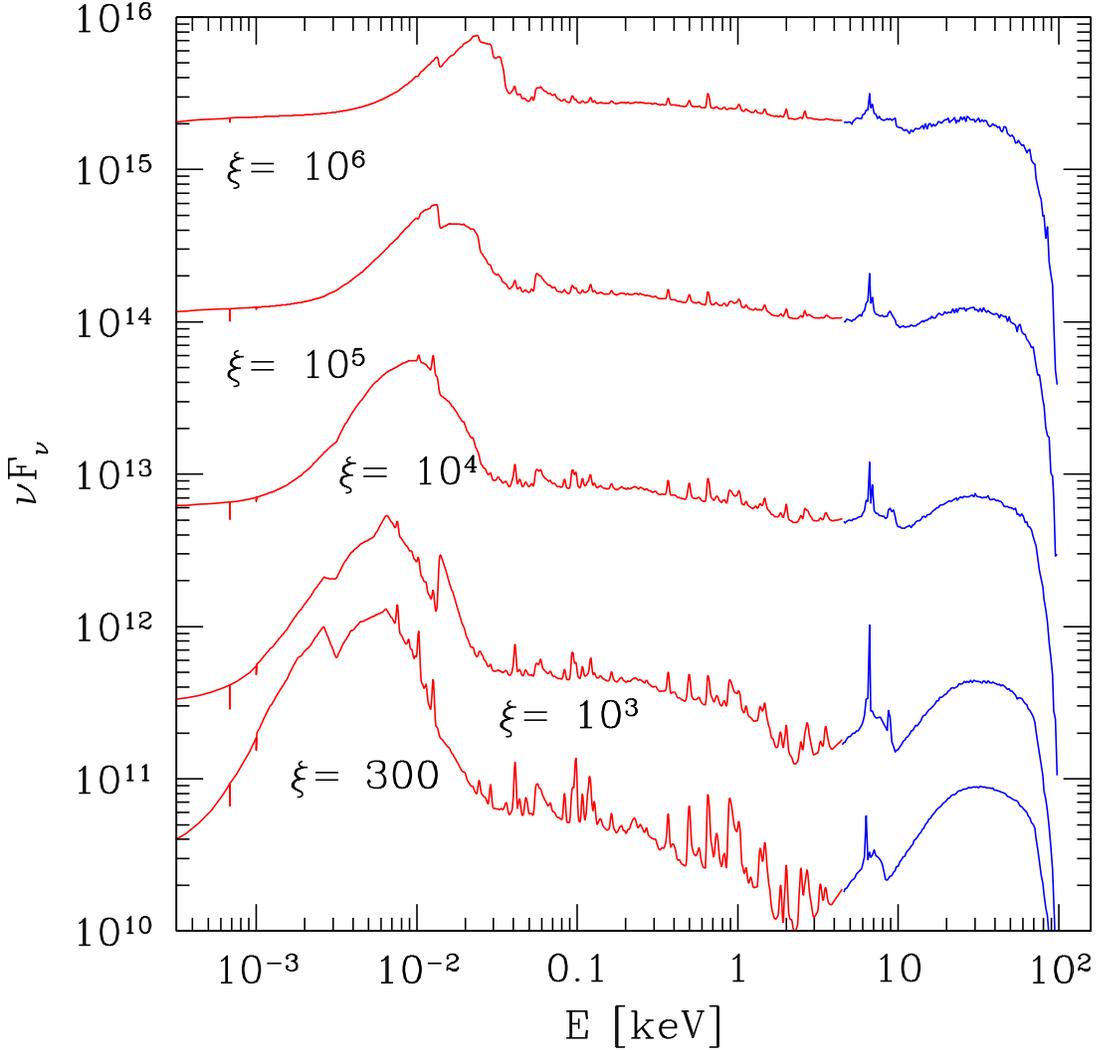}
\caption{The reflection component 
from a constant pressure medium for $\xi = 300,
10^3, 10^4, 10^5$ and $10^6$; $n(0)=10^{11}{\rm cm}^{-3}, N_{\rm H} = 3 
\times 10^{25}{\rm cm}^{-2}$. Spectral resolution: 30.}
\label{fig:spec1}
\end{figure*}

The reflection spectra are given in Fig.~\ref{fig:spec1}. The thermalized
fraction of the incident radiation decreases with increasing 
$\xi$ because the reflected fraction increases, due to the increased value 
of the temperature and of the optical depth of the hot layer, and 
the temperature of the UV
bump increases because the temperature of deep layers increases. 
The Big Blue Bump due to the thermalized radiation is
rather narrow and {\it it never extends to the observed soft X-ray band\/}. 
The same effect occurs in the case of a medium in hydrostatic equilibrium
(NKK, \Agata et al.\ 2002). In the 
far UV band the Lyman edge can be seen in emission (for low values of $\xi$)
or weakly in absorption (for $\xi \geq 10^5$). This is simply due to the 
fact that the underlying continuum becomes more intense in this band as 
$\xi$ increases.  
Soft X-ray lines superimposed on a reflected continuum 
are clearly visible in all cases. The continuum itself clearly shows the
signatures of the ionized reflection since it is relatively flat on 
$\nu F_{\rm \nu}$ plot and slightly rising below 2 keV, although not as 
strongly as in the
constant density solutions (cf. Sect.~\ref{sect:constdens}).

\begin{figure}
\epsfysize=10cm\epsfbox[18 200 600 700]{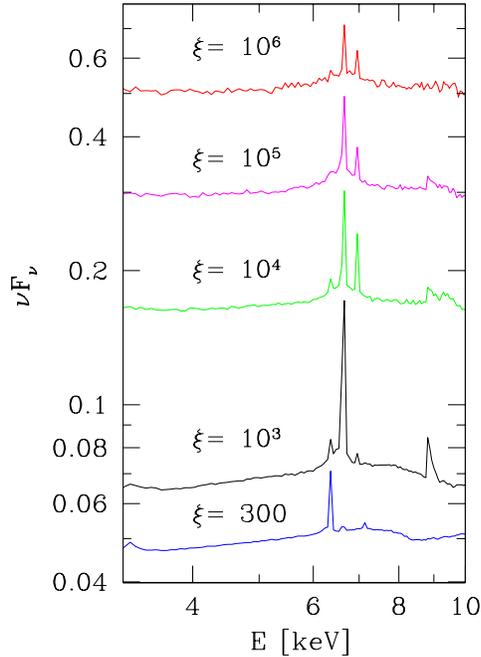}
\caption{The total spectrum (reflection plus primary) in arbitrary units 
in the region of the iron line, computed with the {\sc noar} code 
for a constant pressure medium; $\xi = 300,
10^3, 10^4, 10^5$ and $10^6$; $n(0)=10^{11}{\rm cm}^{-3}, N_{\rm H} = 3 \times 10^{25}{\rm cm}^{-2}$. Resolution: 100.}
\label{fig:kline1}
\end{figure}

The  iron \Ka line in all models consists of several components. 
The neutral component dominates only for $\xi =300$. 
In all
other cases the line comes mostly from ionized iron ions. The helium-like
iron contribution dominates for $\xi = 10^3$, and the role of the hydrogen-like
iron increases with further increase of $\xi$. 
 This is not surprising if
we consider the stratification of iron ion populations  
(see Fig.~\ref{fig:ion1}).
 The details of the spectra in 
this energy band
determined from {\sc noar} are shown in Fig.~\ref{fig:kline1}. 
Equivalent widths
of the line components determined with respect to the total (i.e. reflected plus
incident) radiation flux are given in Table~\ref{tab1}.

The line is broadened by Comptonization and a red tail of the line
develops for large values of the ionization parameter. 
A narrow core,
or rather several narrow components remain, and the effect of Comptonization 
can be seen as an additional
broad underlying component.

\subsection{Comparison with a constant density case}
\label{sect:constdens}

\begin{figure}
\epsfxsize=0.5\textwidth \epsfbox[18 200 600 700]{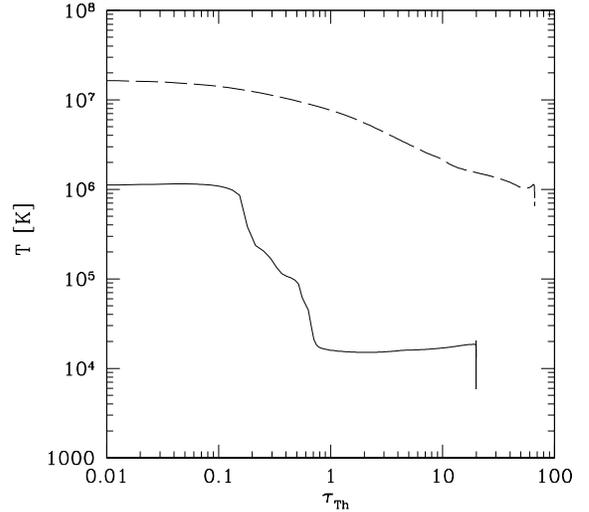}
\caption{The temperature profile in a constant density medium for $\xi =
300 $, $N_{\rm H} = 3 \times 10^{25}{\rm cm}^{-2}$ (solid line), and for
$\xi = 10^5$; $N_{\rm H} = 10^{26}{\rm cm}^{-2}$;  
$n(0)=10^{11}{\rm cm}^{-3}$ (dotted line).}
\label{fig:temp3}
\end{figure}

The reflection from the constant density medium was already broadly 
discussed in the literature 
(Ross \& Fabian 1993, \. Zycki et al.\ 1994, Dumont \& Collin 2001), 
including the Fe \Ka  line properties (Matt et al. 1993, 
\. Zycki \& Czerny 1994).
However, the comparison is much easier if the computations are made with the
same code and for the same sets of corresponding parameters.

Therefore we show two examples, computed for an incident 
power law spectrum, the density equal to $10^{11}$  cm$^{-3}$ 
everywhere in the medium and the
total column density to $3 \times 10^{25}$ cm$^{-2}$ as before.

\begin{figure}
\epsfxsize=0.5\textwidth \epsfbox[18 200 600 700]{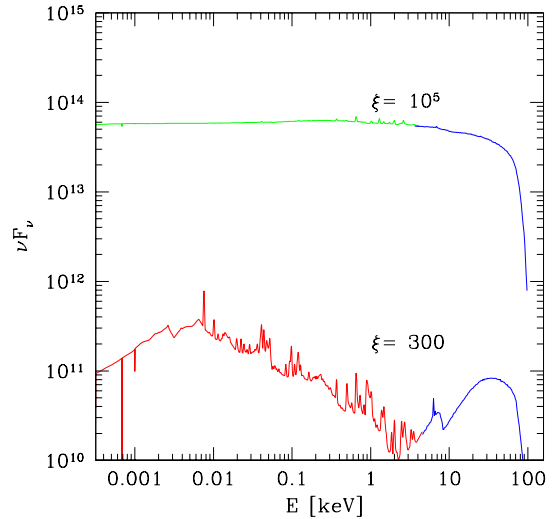}
\caption{The reflection component from a constant density medium, for $\xi =
300$, $N_{\rm H} = 3 \times 10^{25}{\rm cm}^{-2}$ , and for
$\xi = 10^5$; $N_{\rm H} = 10^{26}{\rm cm}^{-2}$;  $n(0)=10^{11}{\rm cm}^{-3}$. 
Resolution: 30.
}
\label{fig:spec2}
\end{figure}

The temperature profile (compare Fig.~\ref{fig:prof1} and \ref{fig:temp3}) 
is much more shallow that in the constant pressure
 case, the heated zone is effectively much broader. The overall spectrum
(see Fig.~\ref{fig:spec2})
varies much more strongly with the incident flux than in the constant pressure case, and displays a much broader ``Big Blue Bump'', due to the presence of
layers with intermediate temperature. This strong, almost a power law, tail 
extends up to 2 keV thus forming a strong soft X-ray excess.  

Despite the more shallow temperature profile the iron line in the constant 
density
case is usually dominated by a single component. This is due to the fact that
the zone where the temperature traces the ionization state is Thomson thick 
and the line is effectively
dominated by the ionization state of the outer layers thus giving a
cold line for low $\xi$, and 6.7 - 6.9 keV lines for higher irradiation flux.
Any particular ionization stage of the iron can be modeled by adjusting the
incident flux, which is not the case either in constant pressure 
computations or in solutions in hydrostatic equilibrium, as noticed by NKK. 

\begin{figure}
\epsfysize=10cm \epsfbox[18 200 600 700]{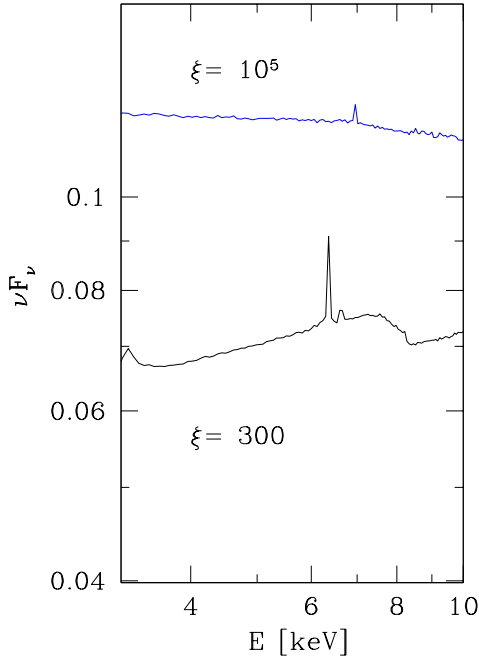}
\caption{The total spectrum (reflection plus primary) in arbitrary units 
in the region of the iron line, computed
 with the
{\sc noar} code 
for a constant density medium; $\xi = 300$, $N_{\rm H} = 3 \times 
10^{25}{\rm cm}^{-2}$ , and for
$\xi = 10^5$; $N_{\rm H} = 10^{26}{\rm cm}^{-2}$;  $n(0)=10^{11}{\rm cm}^{-3}$. 
Resolution: 100.}
\label{fig:kline2}
\end{figure}

In the case of high ionization parameter ($\xi = 10^5$) the line again is 
broadened by  Compton downscattering (see Fig.~\ref{fig:kline2}) 
 and the effect is stronger
in the constant density medium than in the constant pressure case (see also 
Abrassart \& Dumont 2000 and Abrassart 2000). In the case of low $\xi$
the line is dominated by the narrow cold component but we also see
a strong absorption edge. Such a feature is only barely seen from the constant 
pressure medium.

\subsection{Comparison with hydrostatic equilibrium}

Two examples of reprocessing by an irradiated accretion disks were 
presented by \Agata et al.\ (2002). The parameterization of such models is
through the accretion rate, and the pressure increases towards the disk equatorial 
plane. However, we can compare two models with similar total pressure at 
the surface.

In the model computed for a black hole mass $M = 10^8 M_{\odot}$,
an accretion rate $\dot m = 0.03 {\dot M}_{\rm Edd}$ and a distance
$ 10 R_{\rm Schw}$, the gas pressure is of the order of $2 \times 10^3$ 
g cm$^{-1}$ s$^{-2}$ in the surface layers, so it approximately corresponds  to
our case $\xi = 10^4$ of constant pressure models.

The surface temperature is 
slightly lower in the constant pressure model
since the slope of the incident radiation is different 
in the two models ($\alpha = 0.9$ was adopted in the hydrostatic equilibrium 
model, while $\alpha = 1.0$ in the present computations). The overall shape of the 
reflected spectrum in X-ray band is similar, apart from the effect caused
by the difference in the adopted incident flux in the two computations. 
The iron \Ka line in both
cases shows all three components (6.4 keV, 6.7 keV and 6.9 keV), but the 
neutral component is relatively stronger in the case of hydrostatic 
equilibrium. 
The Big Blue Bump peak is also relatively stronger in 
the case of hydrostatic equilibrium. This is partially due to an additional 
energy generation in the accretion disk itself 
and partially due to a much thicker
hot skin in the case of constant pressure medium ($\sim 0.7$ instead of 
$\sim 0.1$) which increases the reflected fraction at the expense of the 
reprocessed one. Increased Thomson thickness of the radiatively heated 
zone
accounts also for the difference in the value of the equivalent width of the 
iron \Ka line -- the line from the constant pressure model is by 50\% 
weaker than in the hydrostatic equilibrium case (66 eV vs.\ 92 eV).

Similar differences are seen between the irradiated disk model  
$\dot m = 0.3$ (hydrostatic equilibrium) and the (constant pressure) model with
$\xi = 10^5$, both having a total surface
pressure of order of $2 \times 10^4$. Also in this case the skin is thicker 
in the constant pressure case ($\tau_{\rm Th} = 1$ vs.\ $\tau_{\rm Th} = 0.3$),
and consequently the equivalent width of the 
Fe \Ka line is lower (80 eV in hydrostatic equilibrium case and 53 eV in 
constant pressure case), and the dominance of the 6.7 keV component in the 
line profile is much stronger in the constant pressure solution.

\subsection{The effect of the spectral shape of incident radiation}

\subsubsection{Power law slope}

The slope of the incident radiation is important not only because of its 
contribution to the observed spectrum. It also modifies the spectral shape 
of the reflected component by affecting the value of the Compton temperature 
and the thickness of the hot layer. 

In constant density computations the incident slope 
determines the reflected slope in the soft part of the spectrum 
(see Fig. 6 of Czerny \&
Dumont 1998) -- a steeper incident flux results in a proportionally steeper 
reflected component. The properties of the iron line do not change 
dramatically.

A medium in hydrostatic equilibrium reacts differently to such 
changes (NKK):
a harder spectrum results in a completely ionized outer skin and 
 a relatively faint
but neutral
iron line produced in the deeper un-ionized zone, while a softer spectrum
gives only  a partially ionized skin and a strong iron line coming from 
ionized iron ions in that region.

\begin{figure*}
\epsfxsize=\textwidth \epsfbox[18 200 600 700]{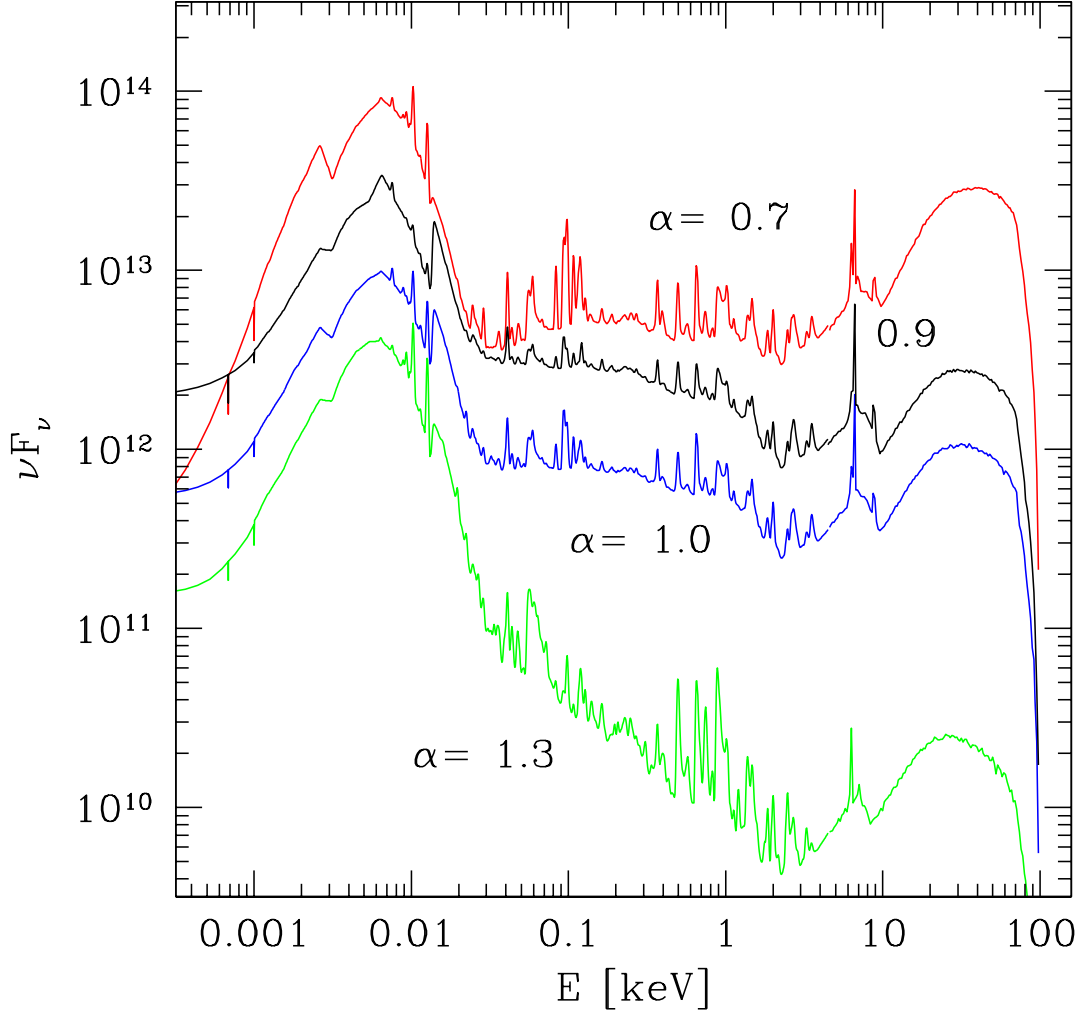}
\caption{The dependence of the reflection component 
from a constant pressure medium on the slope  
of the incident radiation
spectrum $\alpha$ equal 0.7, 0.9, 1.0 and 1.3, for $\xi = 10^3$; 
$n(0)=10^{11}{\rm cm}^{-3}, N_{\rm H} = 3 \times 10^{25}{\rm cm}^{-2}$. Spectra are shifted for better illustration. Resolution: 30.}
\label{fig:specalpha}
\end{figure*}

The reaction of a medium at constant pressure is intermediate between these two
cases (see Fig.~\ref{fig:specalpha}). When the radiation spectrum is very hard ($\alpha = 0.7$) the 
dominant contribution to the line comes indeed from the deep layers 
and the 6.4 keV
component dominates. Intermediate slopes enhance the contribution from
helium-like iron and both line components (6.4 and 6.7 keV) are clearly
seen. However, when the spectrum is very soft ($\alpha = 1.3$), a cold iron line
dominates again since the gas is not heated to a high enough temperature 
(even at the surface) for highly ionized iron to contribute.

\subsubsection{The spectrum of Laor et al.\ (1997)}

A hot (thermal) plasma is a source of a power law radiation which is formed due
to Compton upscattering of the soft photons by energetic electrons. A 
cold layer thermalizes 
a fraction of the radiation but nevertheless, in the plane-parallel
geometry, the X-ray primary emission dominates the total spectrum. This is not
a good representation of high accretion rate objects like quasars or NLS1 
galaxies (where the soft thermal component dominates), which means that 
either the geometry is different, or most of the energy is dissipated within 
the  cold phase at larger distances from the black hole.
%(e.g. accretion disk). 
In this latter case the considered innermost cold 
layer will see not only the hard X-ray source but it will be also affected
by the mean radiation field. 

To illustrate this situation we consider the case of a layer irradiated by a 
source with a spectral distribution given by the mean spectral shape
of quasars, as determined by Laor et al.\ (1997) from a large quasar
sample. This radiation field is much softer than our assumed power law,
since it is strongly dominated by 
the Big Blue Bump, and the corresponding Inverse Compton temperature is lower 
than for a power law spectrum with 
$\alpha = 1.0$ ($2.4 \times 10^6$ K and $1.9 \times 10^7$ K, correspondingly).

\begin{figure}
\epsfxsize=0.5\textwidth \epsfbox[18 200 600 700]{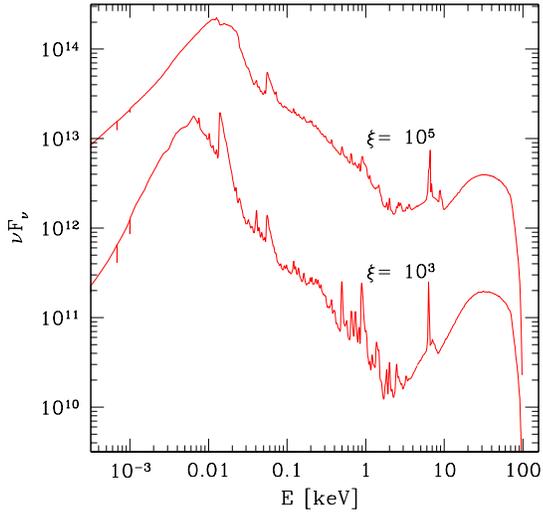}
\caption{The reflection component 
from  a constant pressure medium for the incident radiation spectrum given by 
Laor et al.\ (1997), for $\xi =
10^3$ and $10^5$. Parameters: $n(0)=10^{11}{\rm cm}^{-3}, 
N_{\rm H} = 3 \times 10^{25}{\rm cm}^{-2}$. Resolution: 30.}
\label{fig:spec4}
\end{figure}

The reflected spectrum in the case of a constant pressure medium 
is shown in Fig.~\ref{fig:spec4}. It traces to some 
extent the incident spectrum, giving a strong soft X-ray excess, unlike in the
case
of an incident power law spectrum. The overall shape is much more similar to 
the case of constant density computations, with a power law spectrum. A strong
soft X-ray excess is clearly seen below 2 keV. 

\begin{figure}
\epsfysize= 10cm\epsfbox[18 200 600 700]{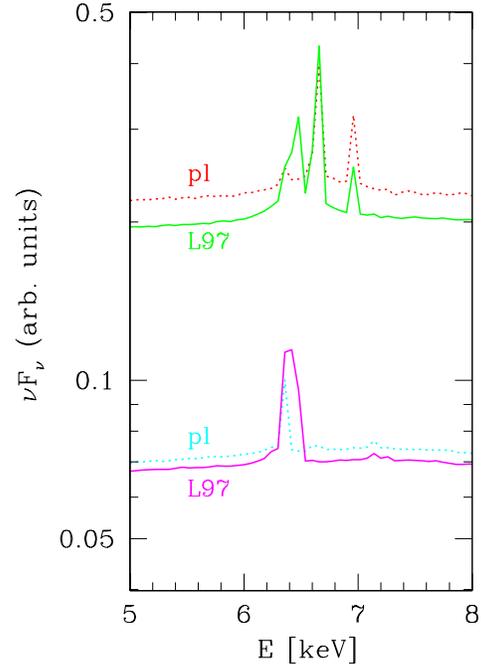}
\caption{The total spectrum (reflected plus primary) in the region of the iron
\Ka line for two sets of models, with similar 
normalization of the incident flux at 2 keV but two shapes of the incident 
radiation spectra. First set: power law spectrum with $\xi = 300$ and
the Laor et al.\ (1997) model with $\xi = 10^3$. 
Second set:  power law spectrum with $\xi = 10^4$ and the Laor et al.\ (1997) 
spectrum with $\xi = 10^5$ from a constant pressure medium. 
Other parameters: $n(0)=10^{11}{\rm cm}^{-3}, 
N_{\rm H} = 3 \times 10^{25}{\rm cm}^{-2}$. Resolution: 100.}
\label{fig:spec8}
\end{figure}

In the \Ka line profile a strong contribution from 6.4 keV component 
is always seen, although
for $\xi \leq 10^5$,  the 6.7 keV component is slightly stronger 
(see Fig.~\ref{fig:spec8}).  A model with a power law incident 
radiation and a comparable hard X-ray flux does not show such a strong neutral
component. 

\subsection{Total column density}

Our computations were deliberately made for a Thomson thick medium. However,
an interesting question is, how strong the effect of the column density is
for a moderately thick medium, and what optical thickness of the medium is 
needed to thermalize the incident radiation.

In the case of a constant density medium the photons penetrate quite deeply into
the medium if the irradiation is strong. We saw in Fig.~\ref{fig:temp3} that
for $\xi = 10^5$ the temperature of the back side was still quite high even for the
very large column density adopted in this computation 
($10^{26}$  cm$^{-2}$).

\begin{figure}
\epsfxsize=0.5\textwidth \epsfbox[18 200 600 700]{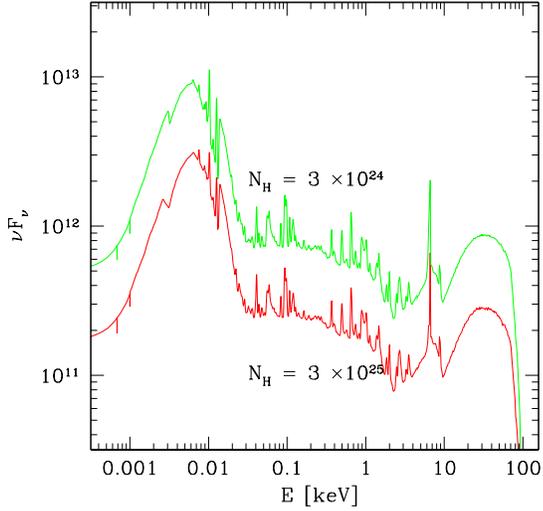}
\caption{The reflection component 
from a constant pressure medium for $\xi = 10^3$; $n(0)=10^{11}$ cm$^{-3}$, 
and two values of the column density:
$ N_{\rm H} = 3 \times 10^{25}{\rm cm}^{-2}$ and $ N_{\rm H} = 3 \times 10^{24}{\rm cm}^
{-2}$. The spectra are shifted for a better illustration. Resolution: 30.}
\label{fig:spec5}
\end{figure}

However, in constant pressure medium the thermalization is much more efficient
for the same Thomson optical depth due to the fact that the density 
is much higher in the cold region, and therefore collisional cooling 
is much more efficient. We give two
examples of the reflection spectra calculated for a moderate value of the
ionization 
parameter ($\xi = 10^3$) and two values of the column density
(corresponding to $\tau_{\rm Th} = 20$ and $\tau_{\rm Th} = 2$; 
Fig.~\ref{fig:spec5}). 
Even if the total optical thickness of the zone is equal to
2 the temperature at the back side of the layer is low. The
resulting spectra are almost indistinguishable.

\subsection{Density}

\begin{figure}
\epsfxsize=0.5\textwidth \epsfbox[30 10 800 530]{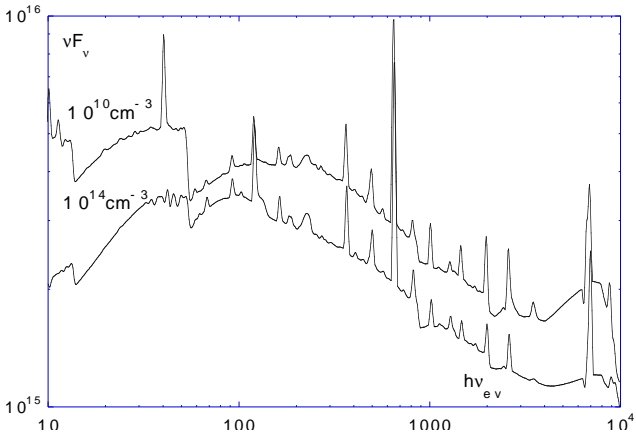}
\caption{The reflection component 
from  a constant density medium for the power law incident radiation spectrum for $\xi =
10^3$ and two values of the number density; $n(0)=10^{10}{\rm cm}^{-3}$ and $n(0)=10^{14}{\rm cm}^{-3}$; $ N_{\rm H} = 3 \times 10^{25}{\rm cm}^{-2}$. 
The spectra are shifted for a better illustration. Resolution: 30.}
\label{fig:spec6b}
\end{figure}

Parameterization of the incident flux with the ionization parameter $\xi$ has
the advantage of reducing significantly the dependence on the medium density.
In constant density computations this approach was particularly succesfull and 
the reflection spectra do not depend significantly on the density in 
a broad range of parameters (from $n = 10^{10}{\rm cm}^{-3}$ up to 
$n = 10^{14} {\rm cm}^{-3}$). Some differences are seen in the low energy band
(IR and optical) due to the influence of free-free emissivity and
the reflection hump above $\approx 3$ keV is more prominent in the low 
density case (Fig.~\ref{fig:spec6b}).

\begin{figure}
\epsfxsize=0.5\textwidth \epsfbox[18 200 600 700]{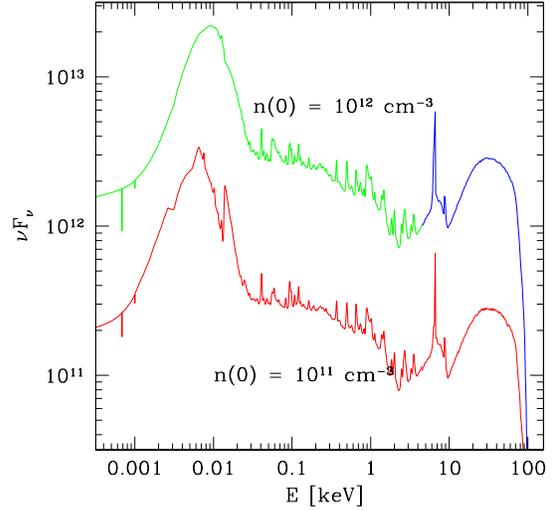}
\caption{The reflection component 
from a constant pressure medium for the power law incident radiation spectrum for $\xi =
10^3$ and two values of the number density; $n(0)=10^{11}{\rm cm}^{-3}$ and $n(0)=10^{12}{\rm cm}^{-3}$; $ N_{\rm H} = 3 \times 10^{25}{\rm cm}^{-2}$.
\label{fig:spec6}}
\end{figure}

Radiation transfer computations performed in constant pressure medium are
much more sensitive to the density itself due to the density gradient. When
the surface density is changed the radiation to gas pressure ratio
changes at the surface and the new density profile as a function of
the optical depth changes, and it is
not just a rescaled version of the old profile by a constant value. This 
induces some differences in the broad band spectrum as well as in 
details of the spectral features (see Fig.~\ref{fig:spec6}).

\subsection{Summary of models}

In order to facilitate the comparison between the various models presented here
and the observed properties of the AGN spectra we determine for those models 
several broadly used quantities. The results are presented in Table~\ref{tab1}
and they include more models than just those presented in the previous
subsections.

Since the computations are made in plane-parallel geometry the final
spectrum consists of the reflected spectrum plus the incident spectrum, without
any additional weight.

We introduce the $\alpha_{\rm ox}$ parameter as the mean slope calculated between
2500 \AA ~ and 2 keV. 
We give the widths of the components of the Fe \Ka line, 
measured with respect to the observed continuum (i.e. reflected plus
incident) 
which result directly from our computations.

Since the observed spectra are usually analyzed using the XSPEC software (and 
the standard spectral components provided with it) we follow other authors
presenting their theoretical results (Done \& Nayakshin 2001, 
Ballantyne et al.\ 2001) 
and we analyze our spectra
as we would analyze an observed spectrum. 

We create artificial data using the {\it fake\/} command in {\sc XSPEC}
for ASCA satellite  and assuming the count rate
at the typical level expected from bright AGN sources with a long exposure.

We fit such data using {\sc pexrav} model (Magdziarz \& Zdziarski 1995) 
to represent the incident power law and 
the reflected continuum, and we add {\sc gaussian} to account for the iron
line. These fits provide us again with the 
normalization of the reflected component, $R_{\rm XSPEC}$, and the photon index 
of the
incident radiation, $\Gamma$. In principle $R_{\rm XSPEC}$ should be 
equal to 1, while $\Gamma$ should be equal to the photon index adopted in our 
model computations. However, 
the shape of the reflected component assumed in  {\sc pexrav} is rather 
different from the one from our computations of radiative transfer
and so both $R_{\rm XSPEC}$  and $\Gamma$ are affected. 
The results for various models are 
summarized in Table~\ref{tab1}. We see that low $\xi$ models show 
a reflection component larger than 1, while reflection from highly 
ionized material give values significantly lower than 1. This effect is much
stronger in the constant density models which are characterized by  a
reflection coefficient of the order of unity 
for  a rather narrow range of values of the 
ionization parameter. Fits to models with the Laor et al.\ (1997) 
incident spectrum gave formally
very large values of the reflection coefficient which was caused by the
presence of the strong soft X-ray excess in those spectra models. We did not 
add any additional component during the fitting to account for this excess
which leads to an artificially steep 'incident radiation' and a large reflection.

We also considered the {\sc XSPEC} {\sc pexriv} model 
(Magdziarz \& Zdziarski 1995) of ionized reflection. 
However, since the shape of the reflection component is different in our 
computations and in  {\sc pexriv}, this  model did not provide a better 
description of 
our model and the returned value of the ionization parameter was usually 
rather low independently from the true value characterizing the model. 
Similar problems were found by Ballantyne et al.\ (2001). 

  \begin{table*}
  \caption{Summary of model properties.
  \label{tab1}}
  \begin{tabular}{rrrrrrrrrrrr}
   \hline
    &       &      &    &       &            &     &               \\ 
No. & $\xi^a$ & incid$^b$&$n^c$ & $N_{\rm H}^{\rm d}$ & $E_{\rm max}^f$&$\alpha_{\rm ox}^g$ & (6.4)&(6.7)&(6.9) & $R_{\rm XSPEC}^h$ & $\Gamma^j$ \\
      &    &  &                  \\
\hline
\hline
\multicolumn{4}{l}{\it constant pressure}\\
\hline
      &     \\
1   &  $3 \times 10^2$  & 1.0  & $1 \times 10^{11}$  & $3 \times 10^{25}$  &  5.21 & 1.31 &18.0 & 3.4 & 0.0 & $1.53 \pm 0.17$ & $2.09 \pm 0.05$\\  
2   & $ 1 \times 10^3$  & 1.0  &$ 1\times 10^{11} $  & $3 \times 10^{25}$  & 6.45  & 1.25 &14.9 &96.7 &5.7  &$ 1.82 \pm 0.10$ & $2.11 \pm 0.03$\\  
3   & $ 1 \times 10^4$  & 1.0  & $1 \times 10^{11}$  & $3\times 10^{25} $  & 8.39  & 1.13 & 4.5 &42.8 &19.2 &$ 1.14 \pm 0.05$ & $2.06 \pm 0.02$\\  
4   & $ 1 \times 10^5$  & 1.0  & $1 \times 10^{11}$  & $3 \times 10^{25}$  & 12.62 & 1.04 &5.8  &37.2 &9.7  &$ 0.88 \pm 0.08$ & $2.05 \pm 0.02$\\
5   & $ 1 \times 10^6$  & 1.0  & $1 \times10^{11} $  &$ 3 \times 10^{25}$  & 22.68 & 1.01 &2.9  &18.8 &8.8  & $0.58 \pm 0.02$ & $2.05 \pm 0.02$ \\
6   & $ 1 \times 10^3$  & 1.0  &$ 1 \times 10^{11}$  &$ 3 \times 10^{24}$  &  6.39 & 1.24 &74.9 &102.6&4.0  & $1.61 \pm 0.09$ & $2.10 \pm 0.03$ \\ 
7   & $ 1 \times 10^3$  & 0.7  &$ 1 \times 10^{11}$  & $3 \times 10^{25}$  &  2.37 & 1.16 &16.0 &78.7 &4.2  &$ 1.71 \pm 0.07$ & $1.81 \pm 0.02$ \\ 
8   & $ 1 \times 10^3$  & 1.3  &$ 1 \times 10^{11}$  &$ 3 \times 10^{25}$  & 5.97  & 1.58 &33.9 &0.0  &0.0  & $0.87 \pm 0.04$ & $2.06 \pm 0.03$ \\
9   & $ 1 \times 10^3$  & 1.0  & $1 \times 10^{12}$  &$ 3\times 10^{25} $  & 19.64 & 1.11 &87.0 & 99.8& 2.1 & $1.83 \pm 0.03$ & $2.11 \pm 0.02$ \\
10  & $ 1 \times 10^5$  &Laor  & $1 \times 10^{11}$  &$ 3\times 10^{25} $  & 11.30 & 1.54 &70.0 &91.1 &13.0 &$17.21 \pm 0.45$ & $2.77 \pm 0.05$ \\
11  & $ 1 \times 10^3$  & Laor & $1 \times 10^{11}$  &$ 3 \times 10^{25}$  &  6.50 & 1.66 &114.5&0.0  &0.0  &$17.95 \pm 0.55$ & $2.76 \pm 0.05$ \\
\hline
\hline
\multicolumn{4}{l}{\it constant density}\\
\hline
12  & $ 3 \times 10^2$  & 1.0  & $1 \times 10^{11}$  &$ 3 \times 10^{25}$  &  4.97 & 1.15 &19.9 &4.0  &0.0& $ 2.19 \pm 0.18$ & $2.12 \pm 0.01$ \\
13  & $ 1 \times 10^5$  & 1.0  &$ 1 \times 10^{11}$  & $1 \times 10^{26}$  &368.6  & 0.99 &0.0  &0.0  &1.0& $ 0.15 \pm 0.06$ & $2.02 \pm 0.01 $\\
14  & $ 1 \times 10^3$  & 1.0  & $1 \times 10^{12}$  &$ 3 \times 10^{25}$  & 10.4  & 1.18 &0.0  &105.9&4.0& $ 1.32 \pm 0.08$ & $2.03 \pm 0.03$ \\
15  &$  3\times 10^2 $  & Laor & $1 \times 10^{14}$  &$ 3 \times 10^{25}$  & 9.86  & 1.55 &89.8 &0.0  &0.0 &$11.13 \pm 0.01$ & $2.78 \pm 0.02 $\\
16  &$  2 \times 10^6$  & Laor &$ 1 \times 10^{11}$  &$ 3\times 10^{24} $  &11.11  & 1.50 &0.0  &0.0  & 7.1&$14.34 \pm 0.01$ & $2.69 \pm 0.02$ \\
\hline
  \end{tabular}

$^a$ - ionization parameter

$^b$ - incident spectrum shape (either AGN mean spectrum after Laor et al.\ (1997), or a power law with an energy index ($\alpha$)

$^c$ - number density at the surface

$^d$ - total column density of the medium

$^f$ - energy at which the spectrum has its maximum at the EUVE range in [eV]

$^g$ - effective spectral index measured between 2500 \AA ¨ and 2 keV

$^h$ -  amount of reflection as determined from {\sc pexrav} XSPEC model

$^j$ -  X-ray slope as determined from {\sc pexrav} XSPEC model

% fits were done assuming gaussian line at reasonable energy with width <0.5 and cut-off 100 keV, for
% ASCA data

  \end{table*}

\section{Discussion}
\label{sec:dis}

X-ray reprocessing in plane-parallel geometry under the assumption of the 
constant density has been studied in a number of papers 
(Ross \& Fabian 1993, \. Zycki et al.\ 1994, DAC; see Dumont \& Collin 2001).
Similar computations were also performed for a stratified medium in 
hydrostatic equilibrium (Raymond 1993, NKK, Ballantyne et al.\ 2001, 
\Agata et al.\ 2002).

In the present paper we consider an intermediate case: the medium under 
constant pressure. It may represent a strongly flattened distribution of
clumpy accretion flow.

Our model, as most of the constant density or hydrostatic models, is 
clearly oversimplified since
we assume that the entire flow is characterized by a single value of the 
ionization parameter, which may not be the case in  a real accretion flow. We 
also did not include here the dissipation which may take place in the cold 
phase of  an accretion flow and consequently enhance the contribution of the 
Big Blue Bump to the overall spectrum. This last effect is only simulated to 
some extent 
by considering the shape of the incident radiation flux in the 
form of the mean quasar spectrum as determined by Laor et al.\ (1997).

Nevertheless, the results illustrate certain important trends which should
help to analyse the conditions in the emitting plasma.

Below we compare the results obtained from the constant pressure medium with 
the observed trends in the broad band spectra of AGN.

\subsection{Observed $\alpha_{\rm ox}$ properties in quasar and Seyfert 
galaxies samples}

The observed broad band spectral index $\alpha_{\rm ox}$ measured between 
2500 \AA ~ and 2 keV was shown for a sample of Seyfert galaxies by 
Walter \& Fink (1993) and
for quasars by Green (1996). Both samples showed a large dispersion of this 
quantity, which could be partly explained by reddening
(e.g. Wilkes et al.\ 1999, Laor
\& Brandt 2002). The average
values are about 1.3 for Seyferts and 1.5 for quasars, with most Seyferts
having values between 1 and 1.6, and most quasars having values between 1.2 
and 1.8. 

Our constant pressure models with an incident power law radiation slope equal 
to
1 have on average somewhat lower values of the $\alpha_{\rm ox}$, around 1.2,
with typical values between 1 and 1.3. 
Since
the observed power law index in Seyfert galaxies is on average not steeper 
than the value adopted in our computations it shows that either there is some
dissipation within the cold medium which
slightly contributes to the Big Blue Bump, or the plane-parallel geometry
adopted in the computations is not correct (for spherical accretion models see
Collin-Souffrin et al.\ 1996, Czerny \& Dumont 1998, and Malzac 2001).  
The effect 
is even stronger in the case of quasars. Not surprisingly, the models with an
incident flux based
on averaged quasar spectra reproduce the observed values of $\alpha_{\rm ox}$ for
those models.

Hydrostatic equilibrium models give for $\alpha_{\rm ox}$ 
results comparable to the constant pressure models (see \Agata et al.\ 2002), with values slightly
lower than seen in Seyfert galaxies, and lower than in quasars. Constant density
models give still lower values ($\sim 1$) and a bigger discrepancy between the
simple prediction and the data.

\subsection{Amplitude of the reflection component}

The observed range of the reflection 
amplitudes for Seyfert galaxies 
($\sim 0.3 - 1.5$;
Zdziarski et al.\ 1999) is rougly consistent with the values obtained from
constant pressure models (0.6 -- 1.8). 
Models with constant density easily produce values
of the amplitude of reflection either much larger or much lower than observed.

However, such a simple model does not explain the observed correlation 
between the slope of the incident radiation and the amount of reflection. 
The observational data show a strong increase of the amount of reflection with 
the steepening of the incident radiation. Our set of models shows an
opposite trend: a model with a steep (soft) incident radiation flux 
($\alpha = 1.3$) shows a fitted amount of reflection much lower than 
a model with a flat (hard) incident radiation ($\alpha = 0.7$). However, our 
computations were done for a single value of the ionization parameter 
$\xi = 10^3$ which might affect the results. We could not repeat the same 
computations for a practically neutral medium since for the moment 
our codes do not allow to
solve the coupled equations of radiative transfer and energy balance 
if the temperature is too low. 

Hydrostatic equilibrium models give lower values of the reflection component
($\sim 0.5$ for the incident spectrum slope$\alpha \sim 1$) than constant 
pressure medium, and they show a clear correlation between the slope of the
incident radiation and the amount of reflection, although the range of values
is somewhat smaller than observed  (Done \& Nayakshin 2001).

Constant density models exhibit the biggest range in the amount of reflection
(from 0 to 2), rather too large in comparison with the data.

\subsection{The Fe \Ka line diagnostics}

The structure and the broadening of the iron line seen in the data is still 
under discussion and the situation is not clear. The broad line profile seen in 
the ASCA data of the source MCG-6-15-30 was confirmed in subsequent 
observations. However, the shape of the iron line in a typical Seyfert galaxy
may be narrower than inferred initially (compare the results of Nandra 
et al.\ 1997 and Lubi\' nski \& Zdziarski 2001 based on ASCA data). 
Recent {\it Chandra\/} data point
towards a very narrow neutral component 
(e.g.\ Yaqoob et al.\ 2001a, Sambruna et al.\ 2001, Kaspi et al.\ 2001) 
but it may be 
superimposed on a broader component originating closer to the central object 
(Yaqoob et al.\ 2001b). The line is therefore most probably multicomponent
 but the components have different shapes and originate in different places. 
 
Our analysis show that the iron line originating from a single region is 
expected to be multicomponent if the medium is at constant pressure when the
slope of the incident radiation spectrum 
is $\alpha \sim 1$, and the medium is not 
neutral (i.e. the ionization 
parameter larger than $\sim 200$). 
The multicomponent structure originates in the
same medium, so if kinematical broadening is essential, it should broaden 
all components similarly. 

A similar multicomponent line is present in the spectra of the hydrostatic models
of \Agata et al.\ (2002). Computations of NKK rather show a single neutral 
component for $\alpha \sim 1$.

A single component, either neutral or ionized, is instead characteristic
for a constant density medium. 

There is some level of intrinsic broadening of 
the line profile due to Comptonization within the reprocessing medium, as
suggested by Czerny et al. (1991) but the effect is not very 
strong in the case of a constant pressure medium or a medium in hydrostatic 
equilibrium. A constant density medium broaden the line more effectively, due to 
the larger optical depth of the hot skin. 

At present, the observations are not accurate 
enough to differentiate between the constant pressure and the constant 
density medium, if the narrow neutral component comes from the outer parts of 
the accretion flow (BLR and/or dusty torus) and the broad component is barely
seen and not well constrained. The study of variability could help to locate 
the emission regions of the different components, and subsequently help to 
find an answer to the question of hydrostatic equilibrium.

%\section{Conclusions}

%???

\bigskip

\begin{acknowledgements}

Part of this work was supported by grant 2P03D01816 of the Polish
State Committee for Scientific Research and by Jumelage/CNRS No.\ 16 
``Astronomie France/Pologne''.
\end{acknowledgements}

\end{document}